\documentclass[12pt]{iopart}
\usepackage{graphicx}
\usepackage{cite}

\usepackage{iopams}  
\begin{document}

\title {Evolution of vortex pinning in the FeSe$_{1-x}$S$_x$ system}

\author{V.A. Vlasenko$^{1}$, A.V. Sadakov$^{1}$, T.A. Romanova$^{1}$, S.U. Gavrilkin$^{1}$, A.V. Dik$^{1}$, O. A. Sobolevskiy$^{1}$, B. I. Massalimov$^{1}$, D.A. Chareev$^{2,3,4,5}$, A.N. Vasiliev$^{4,5}$, E. I. Maltsev$^{1}$,  T. E. Kuzmicheva$^{1}$. }	
	
 \address{$^1$ Ginzburg Center for High Temperature Superconductivity and Quantum Materials, P.N. Lebedev Physical Institute RAS, 119991, Moscow, Russia.\\
$^2$ Institute of Experimental Mineralogy, Russian Academy of Sciences, 142432 Chernogolovka, Russia.\\
$^3$ Kazan Federal University, 18 Kremlyovskaya Str., Kazan, 420008, Russia.\\
$^4$ Ural Federal University, Ekaterinburg, 620002, Russia.\\
$^5$ Faculty of Physics, M.V. Lomonosov Moscow State University, Moscow 119991, Russia. }

\vspace{10pt}

\begin{abstract}
We present a comprehensive study of vortex matter and pinning evolution in the FeSe$_{1-x}$S$_x$ system with various doping degree. The influence of sulphur substitution on vortex pinning and peak effect occurrence is studied. We show that there is a complex interplay among various pinning contributions in the FeSe$_{1-x}$S$_x$ system. Additionally, we study a possible vortex liquid - vortex glass/lattice transition and find an evidence that the vortex liquid - vortex glass phase transition in FeSe has a quasi two- dimensional nature. We investigate the upper critical field behaviour in FeSe$_{1-x}$S$_x$ system, and found that the upper critical field is higher than that predicted by the  Werthamer-Helfand-Hohenberg (WHH) model, whereas its temperature dependence could be fitted within a two-band framework. Finally, a detailed H-T phase diagram is presented. 
\end{abstract}
\maketitle

\section{Introduction}

The vast and multifarious family of iron-based superconductors (IBS) brought us quite a few astonishing phenomena, several fundamental and unanswered so far questions, and a handful of record-breaking superconducting properties. FeSe superconductor having the simplest crystalline structure in the family, remains still one of the most attractive both experimentally and theoretically due to its fascinating diversified physical properties. Bulk FeSe crystal shows rather moderate T$_c$ of 8-9K \cite{Hsu}, however enhanced by chemical doping up to 10.5K for FeSe$_{1-x}$S$_x$ \cite{Abdel}, and to 14K for FeSe$_{1-x}$Te$_x$ \cite{Noji}. Under high pressure T$_c$ reaches 37 K \cite{Medvedev}, whereas electric field gating and chemical intercalation increases T$_c$ up to above 40K \cite{Shiogai}. 
Apart from described above phenomena, IBS in general and FeSe in particular exhibit a number of exciting properties in the area of vortex matter physics, attracting interest in both fundamental investigations and research for practical applications. For instance, vortex glass phase \cite{Blatter},  the peak effect (PE) \cite{Ge}, the second magnetization peak (SMP) effect on the magnetization hysteresis loop (MHL) \cite{Galluzzi1}, and a multiband nature \cite{Aswathy} have been observed in IBS. 

The SMP or the so-called "fish-tail" effect is well-known in conventional and high temperature superconductors. It occurs, when the increase of magnetic field leads to enhanced critical current density (J$_c$). The effect was observed in FeSe$_{1-x}$Te$_x$\cite{Wu,Marco Bonura}, BaFe$_{2-x}$Ni$_x$As$_2$ \cite{Pervakov}, BaFe$_{2-x}$Co$_x$As$_2$ \cite{Shen} single crystals, with it, several models were proposed to describe the effect \cite{Fang}. The vortex glass state was evidenced in the FeSe$_{1-x}$Te$_x$ \cite{Yi Yu}, (Ba,K)Fe$_2$As$_2$\cite{Mak}, BaFe$_{2-x}$Ni$_x$As$_2$ \cite{Eltsev}, and in the oxypnictides SmFeAsO$_{0.85}$\cite{Lee} crystals. However, to date, vortex glass phase transition has not been studied in detail in the FeSe$_{1-x}$S$_x$-based superconductors.  

In this paper, we present a detailed transport and magnetic measurements of the  FeSe$_{1-x}$S$_x$ single crystals within the wide temperature range. We discuss  the possible nature and correlation between the SMP, PE, and the doping level. Additionally, we measured the upper critical field H$^c_{c2}$) in magnetic fields up to 19T. All the  (H$^c_{c2}(T)$ curves obtained could be successfully fitted with a two band model\cite{Gurevich}, being inconsistent with the single band WHH model \cite{WHH}. Our results show that the S doping up to x$\leq$ 0.11 in the multiband FeSe$_{1-x}$S$_x$ superconductors changes insignificantly the effective band structure.

\section{Experimental details}
For our studies, we used single crystals of sulphur-substituted FeSe$_{1-x}$S$_x$ with $x$ =0.04, 0.09, and 0.11. The results obtained were also compared with the data on two pure FeSe single crystals (S1 and S2). The main steps of  FeSe$_{1-x}$S$_x$ crystal growth are described in detail in ref.\cite{Chareev}. A complete characterization of our samples was detailed elsewhere \cite{Abdel}. Magnetization measurements were performed using a Quantum Design PPMS-9. The typical field sweep rate was retained with 100 Oe/s. 
The temperature dependent electronic transport was measured with a four-probe AC method with a current applied parallel to the crystallographic $ab$ plane. High-field measurements were done in 16T and 21T superconducting magnets  (Cryogenic Ltd.) at temperatures down to 0.3K. Tranport I-V measurements for vortex glass scaling experiments were performed on a micro bridges, made with a focused ion beam lithography on Helios NanoLab 660. In Fig.1(a-d) we present the standard four-probe R(T,H) measurements of our FeSe$_{1-x}$S$_x$ samples in magnetic fields up to 19T. We define the temperature of the superconducting transition using the T$^{50\%}_c$=(T$^{onset}_c$+T$^{offset}_c$)/2  criterium as it shown in Fig.1c. 

\begin{figure}[t]\vspace*{3pt}
	\includegraphics [width=1.1 \textwidth]{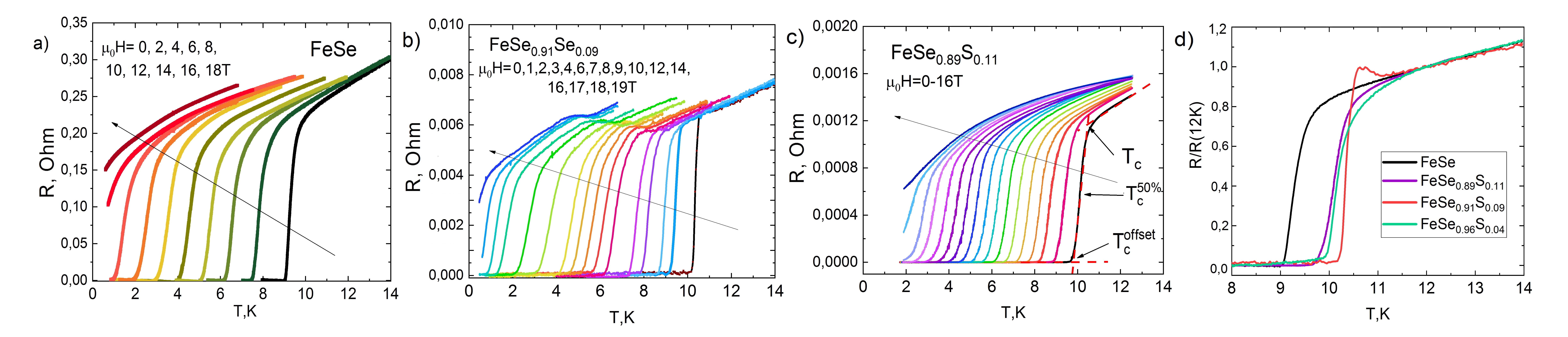}
	\caption{The R(T) measurements of FeSe$_{1-x}$S$_x$ single crystals with a)x=0 b)x=0.04 c) x=0.11 single crystals in  magnetic fields up to 19 T along the c-axis. d): The resistive superconducting transition of FeSe$_{1-x}$S$_x$ single crystal in zero magnetic fields for x=0, 0.04, 0.09, 0.011.}
	
\end{figure}

\section{Vortex pinning mechanism}

\subsection {Critical current}
We investigate vortex pinning  using several models and methods. Fig.2 shows the isothermal magnetization hysteresis loops  with H$\parallel $c  obtained at various temperatures. The symmetry of MHL points to relatively weak surface barriers and a strong bulk pinning. This fact also indicates our samples contain only a negligible amount of magnetic impurities, i.e. all Fe atoms exhibit a compensated magnetic moment\cite{Mahmoud2}. Magnetization curves show a presence of a second peak at low temperatures ( T$\leq$3K ) and H$\parallel $c  for all the samples, whereas the PE develops only in the S substituted samples. It is noteworthy that, SMP appears significantly weaker at 2K in one pure FeSe single crystals, but rather well at 1.45K in the another crystal (S2) from the same batch (see Fig.2a,b). 

\begin{figure}
	\includegraphics[width=1\textwidth]{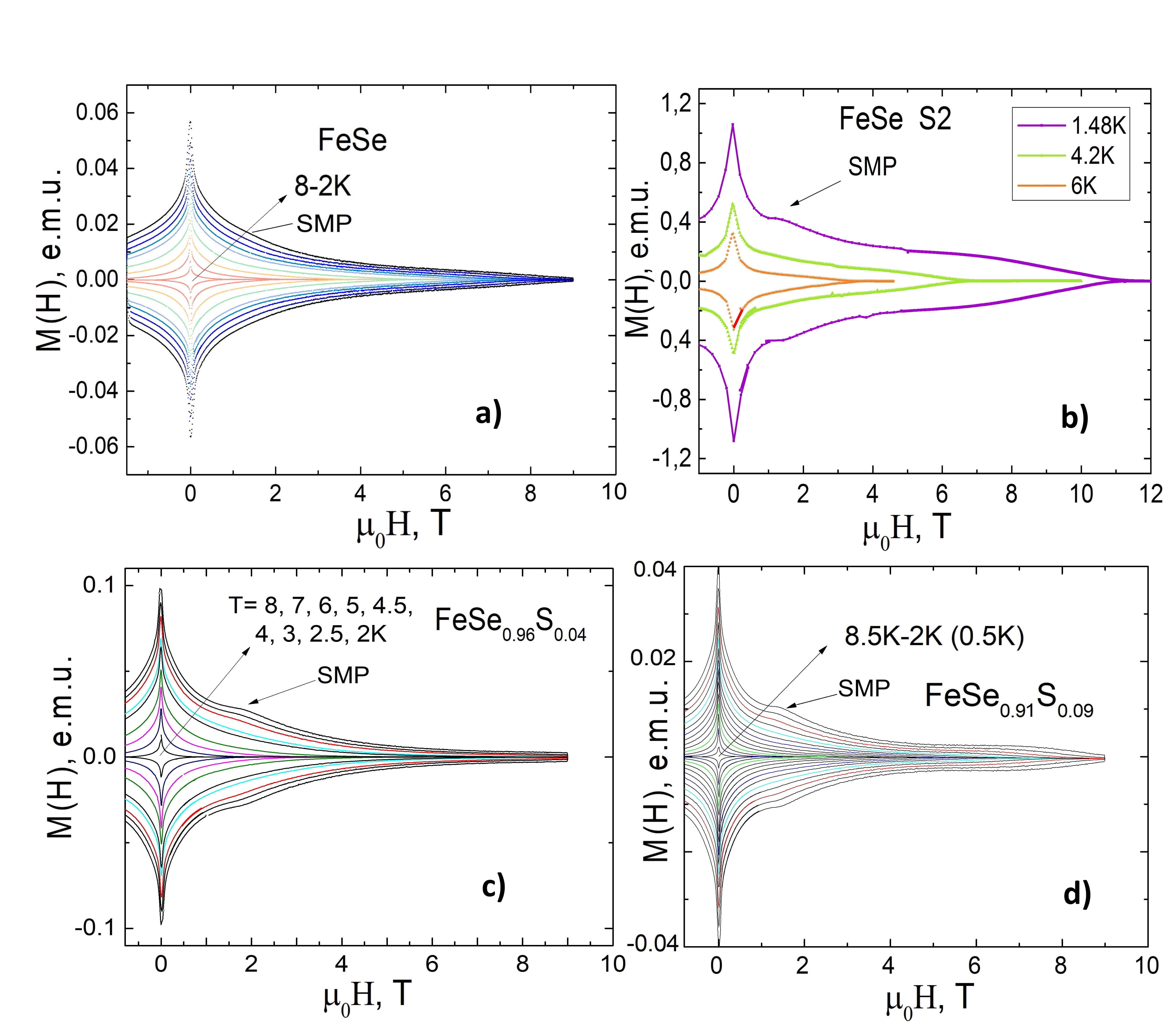}
	\caption{Isothermal magnetization hysteresis loops as a function of magnetic field with H $\parallel $ c  up to 9 T are shown for FeSe$_{1-x}$S$_x$ at various temperatures with x = 0 , 0.04, 0.09, respectively.}
\end{figure}

From the MHL, we obtained the magnetic field dependence of the critical current density J$_c$(H) using Bean critical state model \cite{Bean} at various temperatures (see Fig. 3). In that model  the critical current density for a platelet sample is given by the formula:
\[J_c = 20dM/(a(1 - a/3b)),    (1)\] 
where $M = M_{dn} -M_{up}$, $ M_{dn}$ and $M_{up}$ are the magnetizations measured under decreasing and increasing fields, respectively, $a$ and $b$ ($b > a$) are the dimensions of the crystal surface perpendicular to the applied field. Here M is given in electromagnetic units per cubic centimeter and the resulting J$_c$ is in A/cm$^2$. The determined J$_c$(H) values using Eq.(1) are shown in Fig. 4 in a log-log plot for x = 0, 0.04 and 0.09 crystals.
Habekhorn et al.  \cite{Haberkorn} showed that J$_c$(H) diagram in the log-log scale facilitates distinguishing between several pinning regimes. At low fields, J$_c$ is independent of H (regime I). With the subsequent H increase, we observed a power law J$_c$ $\propto$ H$^{-a}$ behavior (regime II). Regime III exists at rather narrow field range where J$_c$(H) $\propto$ $const$. Finally, regime IV is observed when J$_c$ rapidly decreases with H and tends to zero at H$_{irr}$. The nature of pinning regimes is the following: (I) the low-field part is associated with the single vortex state; (II) the power-law dependence J$_c$ $\propto$ H$^{-a}$ is associated with strong pinning centers; (III) coheres with fishtail effect; and (IV) related to changes in the vortex dynamics. 

\begin{figure}
	\includegraphics[width=1\textwidth]{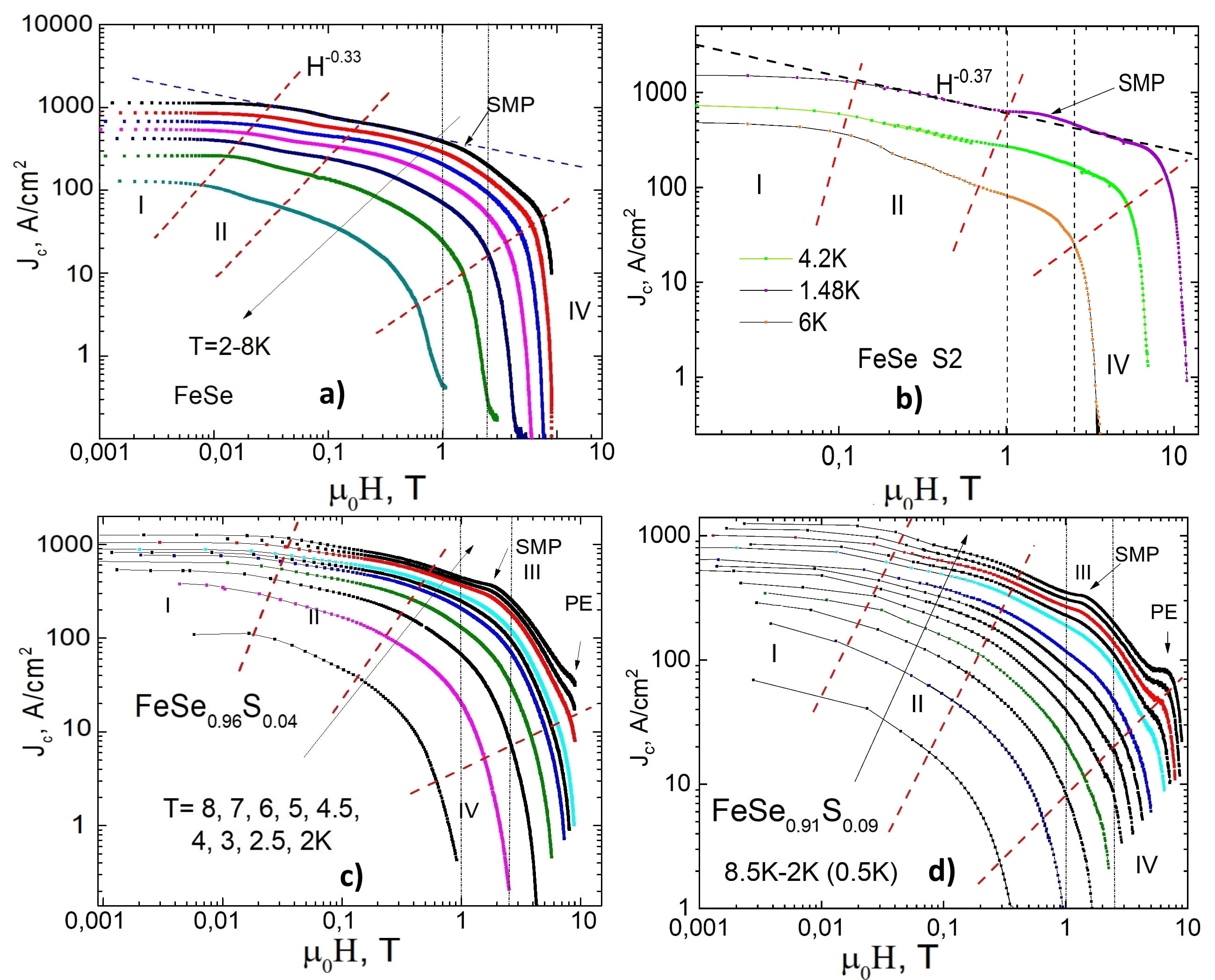}
	\caption{Magnetic field (H) dependence of the critical current density (J$_c$) for the FeSe$_{1-x}$S$_x$ single crystals at various temperatures with x = 0.0 ; 0.04; 0.09. The red dashed lines indicate the start and the end of the different pinning regimes. The vertical lines shows the value of applied magnetic field causing the vortex lattice rearrangement in FeSe according ref.\cite{Putilov}.  }
\end{figure}

All the data for  x=0, 0.04 and 0.09 presented in Fig.3 shows some qualitative similarities. The initial J$_c$(H) $\propto$ $const$ (regime I) behaves at fields about 100-150 Oe. In higher fields we observe that critical current follows a power law J$_c$ $\propto$ $H^{-a}$ up to 1T with  $0.33<a<0.72$ for x=0, $0.18<a<0.59$ for x=0.04, and $0.25<a<0.58$  for x=0.09 crystal. The obtained values of the exponent at higher temperatures are in a good agreement with the theoretical prediction H$^{-5/8}$, which indicates strong vortex pinning \cite{Beek2}. It is worth mentioning that with temperature decrease, the value of  $a$ diminishes. We assume that the ${a}$ value could decrease down to 0.2 due to the presence of columnar-like defects, whereas the intermediate values relate to extended defects and nanoparticles \cite{Haberkorn3}. Taking into account that the sulphur doping suppresses twinning completely only for $x> 0.17$  \cite{nematic_suppress}, twin boundaries  in our samples could act as  extended  defects.  However, rather low $a$ values at low temperatures can not be explained only by the influence of twin boundaries.  More likely, the strong pinning relates to the point-like pinning centers due to the distortion of the crystal lattice. It should be mentioned that, according to our data, S substitution also reveals the peak effect at low temperatures (T$\le3K$). The nature of the SMP, PE, and strong pinning contribution strengthen with temperature decrease will be discussed below. 

\subsection {Vortex pinning force}

In order to investigate the vortex pinning mechanism in more detail, we calculated the pinning force density $F_p$ from the critical current density, and the applied field, using $F_p = B \times J_c$ at various temperatures. As proposed by Dew-Hughes \cite{Dew}, the pinning mechanism does not change with temperature, if the normalized pinning force $f_p = F_p/F_{p max}$ as function of reduced field $h_p= H/H_{irr}$ demonstrates a scaling relation, $f_p(h_p) \propto h^p(1-h)^q$ (here $F_{p max}$ is the maximum pinning force, $p$ and $q$ are the exponents). The values of $p$ and $q$ depend on the defect dimensionality (point, two-dimensional, or bulk), the type of interaction, and the nature of pinning centers. The deviation from the scaling law with  temperature or magnetic field points to the change in vortex-lattice period or the various size of pinning centers. 

\begin{figure}
	\includegraphics[width=1\textwidth]{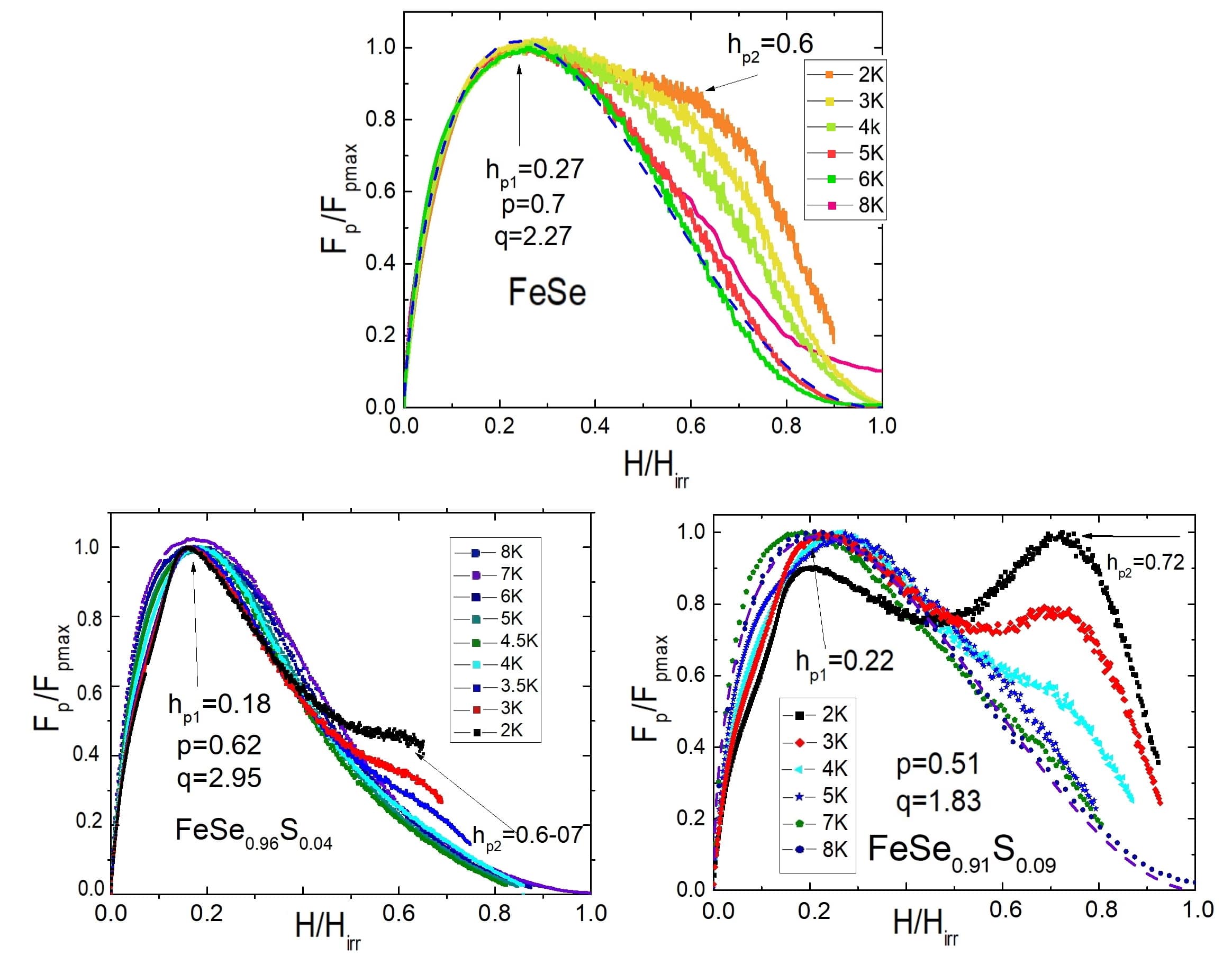}
	\caption{Normalized pinning force as a function of reduced field for FeSe$_{1-x}$S$_x$ :  with x = 0.0; 0.04 and 0.09.}
\end{figure}

In Fig.4  we present the $f_p(h_p)$ data for FeSe$_{1-x}$S$_x$ single crystals at various temperatures. At low temperatures (T$\leq$0.5T$_c$), all the curves deviate from a single asymptotic behavior. For instance, the FeSe$_{1-x}$S$_x$ compounds with x=0.04 and 0.09 tend to change the character of the dependence from a curve with one narrow peak to a curve with two peaks. In contrast, the pure FeSe sample shows the widening of the $f_p(h_p)$ peak with temperature decrease. The first peak in S-doped samples is located around h$_{p1}\approx$0.2; the second one at h$_{p2}\approx$0.65. According to the theory \cite{Dew, Koblischka1}, $F_p$ vs. $h$ curves at h = 0.2 (p = 1/2, q = 2) are characteristic for surface pins, such as grain boundaries or planar pins; h = 0.33 (p = 1, q = 2) $-$ for point pins ($\delta$l pinning). The second maximum of F$_p$(h) at h = 0.67 (p = 2, q = 1) relates to point pins and h = 0.6 (p = 3/2, q = 1) corresponds to surface pins ($\delta$Tc pinning), respectively. With such complex behavior of F$_p$(h) plots  it seems highly problematic to determine uniform dominant pinning mechanism. However, in the sulphur substituted  samples we can distinguish the two main areas of $f_p/h_p$ plot. The f(h) peak at high fields (PE) is attributed to weak and widely spaced pins, while strong closely spaced pins induce the large peak at low fields in S-substituted samples. Thus, in FeSe$_{1-x}$S$_x$ samples, we observe at least two different field and temperature dependent  pinning mechanisms. 

Another significant issue is the influence of  anisotropy on the physical properties of the superconductor, unaccounted in Dew-Huges and Kramer models. According to the work \cite{Eisterer}, a percolation of the current between the grain boundaries was shown for anisotropic superconductors, which leads to a significant shift in h$_{p}$ position to lower values with the anisotropy increase. In case of FeSe$_{1-x}$S$_x$ compounds, the anisotropy of physical parameters is usually less than 2, with a tendency to decrease at low temperatures, thus neglecting the percolation effect. Therefore, despite the above mentioned model limitations \cite{SANDU1}  for superconductors, the Dew-Hughes and Kramer model is applicable for qualitative description of the vortex pinning in IBS compounds with low anisotropy.

\subsection {The origin of SMP and PE }

The origin of SMP has been studied extensively in cuprate superconductors and was attributed to various mechanisms: crossover from elastic to plastic vortex creep \cite{Abulafia}, vortex order-disorder phase transition  \cite{Nishizaki2}, vortex lattice structural phase transition (VL) \cite{Rosenstein}, surface barriers  \cite{Kopylov}, vortex lattice transition from 3D  to 2D \cite{Prozorov2}, etc. The possible origins of SMP suggested by Barilo et al. \cite{Barilo} are as follows: 1) the two-dimensional (2D) character of pinning centers; 2) two different types of pinning centers: normal core and $\delta$Tc; 3) crossover between different pinning mechanisms is induced by external magnetic field and/or temperature. 

The cause and nature of SMP and PE in iron-based FeSe$_{1-x}$S$_x$ single crystals seems to be a rather complicated issue.  According to ref. \cite{Galluzzi}, the SMP in FeSe sample can be caused by their multi-domain and twinning nature (where domain walls or twin boundaries play a role of elongated planar defects). However, in FeSe$_{1-x}$S$_x$ samples twinning diminishes with S concentration increase and vanishes at $x>0.17$ \cite{nematic_suppress},  while SMP increases. Therefore,  twinning cannot be the main cause of the second magnetization peak appearance.  In FeSe(Te) compound the SMP was attributed to order-disorder transition in vortex matter \cite{Miu, Das}. Song et. al.\cite{STM_image} showed a vortex lattice ordering image in different magnetic fields at T=0.4K on an MBE grown film with ~10nm thickness. Moore et. al. \cite{moore2}  made STM vortex images on the FeSe$_{1-x}$S$_x$ samples with the same doping degree. The authors showed clear hexagonal vortex lattices in fields 5-6T at 1.5K, exactly next to the SMP, as compared with our MHL curves (see Fig. 2). Moreover, in a recent work\cite{Putilov}, a vortex  lattice  strong  deformation from an almost hexagonal to  square lattice was observed at applied magnetic fields from 2.5T to 1T. Our MHL data shows that SMP position in all FeSe$_{1-x}$S$_x$ samples is around 1.5T. Therefore, we can conclude that vortex structure rearrangement can be the main reason of SMP in FeSe$_{1-x}$S$_x$ superconductors. 

The nature of peak effect is another interesting issue of the FeSe$_{1-x}$S$_x$ system. Our data clearly shows that PE correlates with S doping degree. It arises at x=0.04 and hugely increases for x=0.09 doping level. This effect occurs in high fields for several reasons: 
-(i) change in vortex structure; 
-(ii) change in pinning regime, i.e. a presence of weak defects, which enhances the pinning force in high fields or effect of caging \cite{Haberkorn};
-(iii) phase inhomogeneity in the bulk of the superconductor, which leads to fluctuations  of the upper critical field ($H_{c2}$ inhomogeneity) or $\kappa$ parameter like in YBCO high-temperature superconductor (HTSC) (weak superconducting region)\cite{Daeumling}. In high fields, normal and superconducting domains, at low temperatures could create additional pinning centers and cause PE to appear, while at higher temperatures phase separation is insignificant for PE to emerge. 
While the broadening of the superconducting transition in R(T,H) measurements with increasing applied magnetic field becomes evident for (iii) case of PE nature, however, more detailed further investigations are needed.

Summarizing the data showing the presence of the SMP, PE, and vortex structure ordering in Figs.2-4, we found an evidence of several  vortex pinning mechanisms  present in FeSe$_{1-x}$S$_x$ system. We found that  PE correlates with S substitution level, possibly caused by the $H_{c2}$ inhomogeneity in high fields. The SMP is observed for all the studied samples and appears due to the magnetic vortex structure deformation.

\subsection{Vortex glass - vortex liquid phase transition}

\begin{figure}
	\includegraphics[width=1\textwidth]{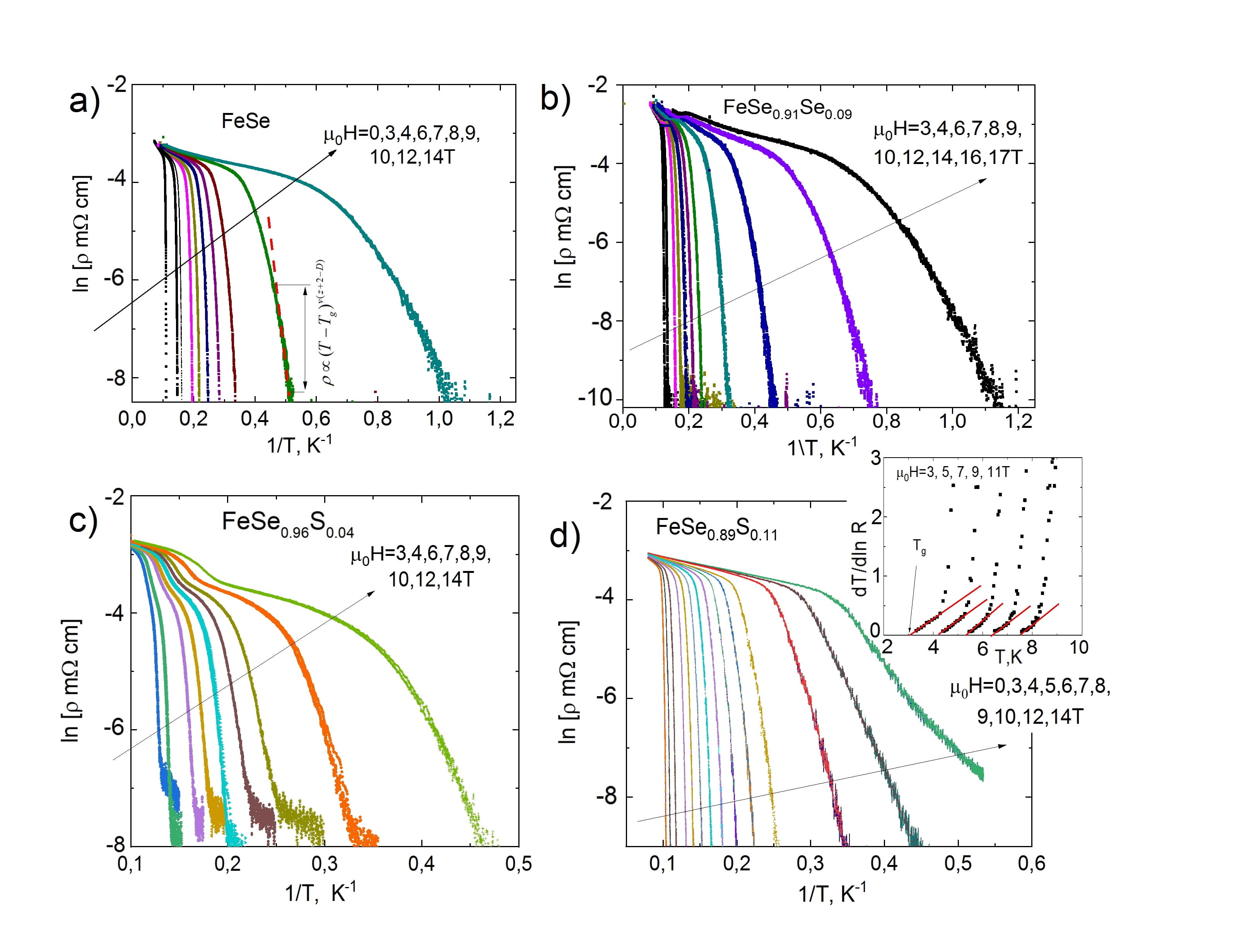}
	\caption{(a-d)  The semi-logarithmic plots of the R(T) data of the FeSe$_{1-x}$S$_x$  single crystals in c-axis magnetic fields up to 14 T. Inset: Inverse logarithmic derivative of resistivity for FeSe$_{0.89}$S$_0._{11}$  at various fields. The dashed lines represent fits with the VG theory. The glass temperature, T$_g$, is determined using the relation $(d\ln\rho /dT)^{-1} =0$.}
	
\end{figure}
\begin{figure}
	\includegraphics [width=1\textwidth]{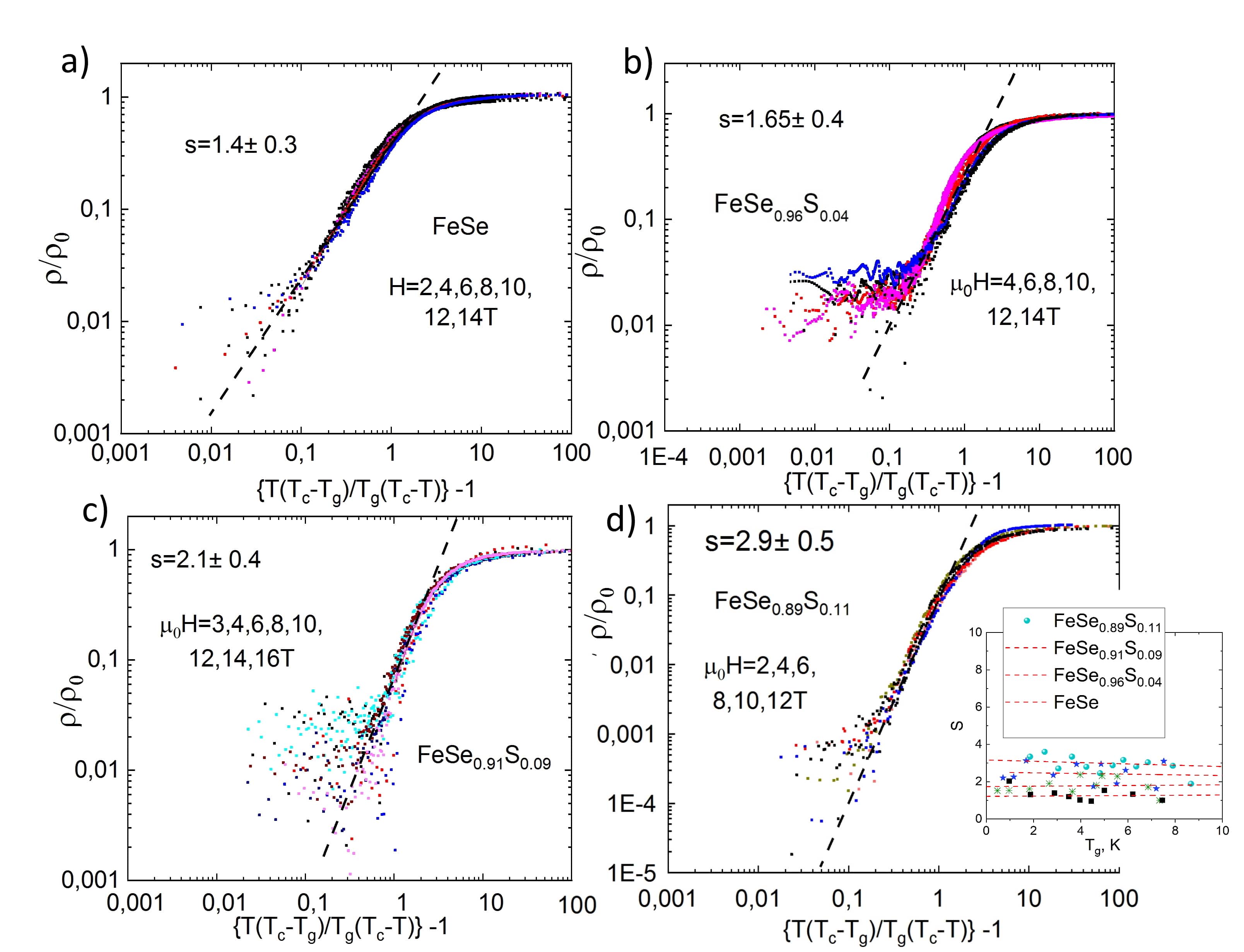}
	\caption{(a-d) Normalized temperature dependence $\rho /\rho _n$  versus $T(T_{c}-T_{g})/(T_{g}(T_{c}-T)-1 )$ in various fields. Inset: The critical exponents $s$ for various fields, obtained within the vortex-glass model for various S substitution. }
	
\end{figure}

It is well known that in the mixed state, thermal fluctuations affect the vortex motion, thus broadening the resistive R(T,H) transition.  The  magnitude of thermal fluctuations is quantified by the Ginzburg number  \cite{Lee} :\[Gi={{10}^{-9}}{{\left[ \frac{{{\kappa }^{4}}{{T}_{c}}[K]{{\gamma }^{2}}}{{{H}_{c2}}(0)[Oe]} \right]}^{2}},   (2)\]  where $\kappa$  is the Ginzburg-Landau parameter,
$H_{c2}(0)$ the upper critical field at zero-temperature, and $\gamma$ the anisotropy ratio between the $ab$-plane and the c-axis coherence length. The typical $Gi$ values are in range $1
0^{-8}-10^{-5}$ for low-$T_c$ conventional
superconductors, $Gi>10^{-2}$ for cuprate superconductors \cite{Kacmarcik}, $Gi\sim 10^{-5}$ for MgB$_2$ \cite{Angst}. The $Gi$ value obtained for IBS are $10^{-4}-10^{-2}$:  Ba(K)-122 ($1-5\times 10^{-4}$) \cite{Kacmarcik},  Nd-1111 ($8\times10^{-3}$), Ba(Co)-122 ($1.7\times 10^{-4}$), Fe-11 ($1.3\times 10^{-3}$) \cite{Putti2}, Sm-1111 ($1.6\times 10^{-2}$)\cite{Welp}. Hence, rather strong  thermal fluctuations in iron based superconductors should produce a variety of vortex phases and vortex dynamics behavior. Here, we estimate the $Gi$ number by using Eq. (2) for FeSe$_{1-x}$S$_x$ single crystals with x=0, 0.04 and 0.009, taking the $H_{c2}$, $T_c$ from our previous work \cite{Abdel}. The calculated  $Gi$ values are $5\times 10^{-3}$, $8\times 10^{-3}$, $3\times 10^{-3}$ for x=0, 0.04, and 0.09, respectively. Thus, we may conclude that thermal fluctuations are not so significant in our system as compared with Cu-based superconductors.  In a recent work \cite{Huan Yang2} the authors claimed that superconducting-fluctuations (SCF)  in bulk FeSe are not strong, with the only very narrow SCF region observed.

In order to understand the effect of sulphur substitution on the vortex properties of our samples, we studied phase transition between vortex glass and vortex liquid. According to the vortex glass (VG) theory \cite{Fisher,Fisher2}, the vortex-solid to vortex-liquid phase transition can be determined from R(T,H) and I(V) measurements. The VG theory predicts that the linear resistivity response should vanish near T$_g$ as $R=(T-T_g)^{(z+2-D)v}$, where D is the sample dimensionality. Here, $v$ and $z$ are the static and the dynamic exponent, respectively.   Therefore, using the equation: $(d\ln\rho/dT)^{-1}=1/s (T-T_g)$ with $s= (z-1)\times v$, we estimate $T_g$ and the critical exponent $s$ from the linear region of the Arrhenius $(d\ln\rho /dT)^{-1}$ vs.  $T$) curves. Figs.5(a-d) show the Arrhenius plots of the R(T) data for FeSe$_{1-x}$S$_x$ single crystals. Using this data, we estimate the $T_g(H)$ values as shown in the inset of Fig.5(d).

The critical exponent $s$ of the temperature dependence was obtained from the best fit of $(d\ln\rho /dT)^{-1}$ vs. $T$ plot (red lines in the inset of Fig.5(d)). Using the resistivity measurements, we obtained 's' values in the range 1.5 $\pm$ 0.5 for pure FeSe, 1.7$\pm$ 0.7, 2.5 $\pm$ 1, 3.2 $\pm$ 1.1 for FeSe$_{0.96}$S$_{0.04}$, FeSe$_{0.91}$S$_{0.09}$ and FeSe$_{0.89}$S$_{0.11}$, respectively. 

Another way to examine the possible VG transition is the scaling law of $R \sim (T-T_g)^ {ν(z-1)}$ near $T_g$.
According to VG theory \cite{Fisher,Fisher2}, resistivity decreases as a power law close to the glass transition temperature ($T_g$): \[\rho ={{\rho }_{n}}{{\left| T/{{T}_{g}}-1 \right|}^{s}},    (3)\] where $\rho_n$ is the characteristic resistivity in the normal state. In the modified vortex glass model proposed by Rydh, Rapp, and Andersson (RRA)\cite{Andersson} scaling expression for resistivity near the $T_g$ changes to:\[\rho ={{\rho }_{n}}{{\left| \frac{T({{T}_{c}}-{{T}_{g}})}{{{T}_{g}}({{T}_{c}}-T)}-1 \right|}^{s}}    (4).\] 
The benefit of RRA model is in use of the two dimensional approach in the B(T) plane in order to determine the $T_g$, instead of taking the usual one dimensional case. In the 2D approach, the R(T) data is expected to collapse onto a single curve more precisely in the $\rho/ \rho_n$ versus $(T(T_c -T_g)/T_g(T_c -T) -1)$ log-log plot, thus facilitating the estimate of $T_g$ and $s$ values.
Fig. 6 (a-d) shows the nice scaling of FeSe$_{1-x}$S$_x$ R(T,H) data based on the RRA approach. The slopes give  $s$ = 1.4 $\pm$ 0.3, 1.65 $\pm$ 0.4, 2.1 $\pm$ 0.4, and 2.9 $\pm$ 0.5 for  FeSe, FeSe$_{0.96}$S$_{0.04}$, FeSe$_{0.91}$S$_{0.09}$, and  FeSe$_{0.89}$S$_{0.11}$ single crystals, respectively. 

\begin{figure}
	\includegraphics [width=1\textwidth]{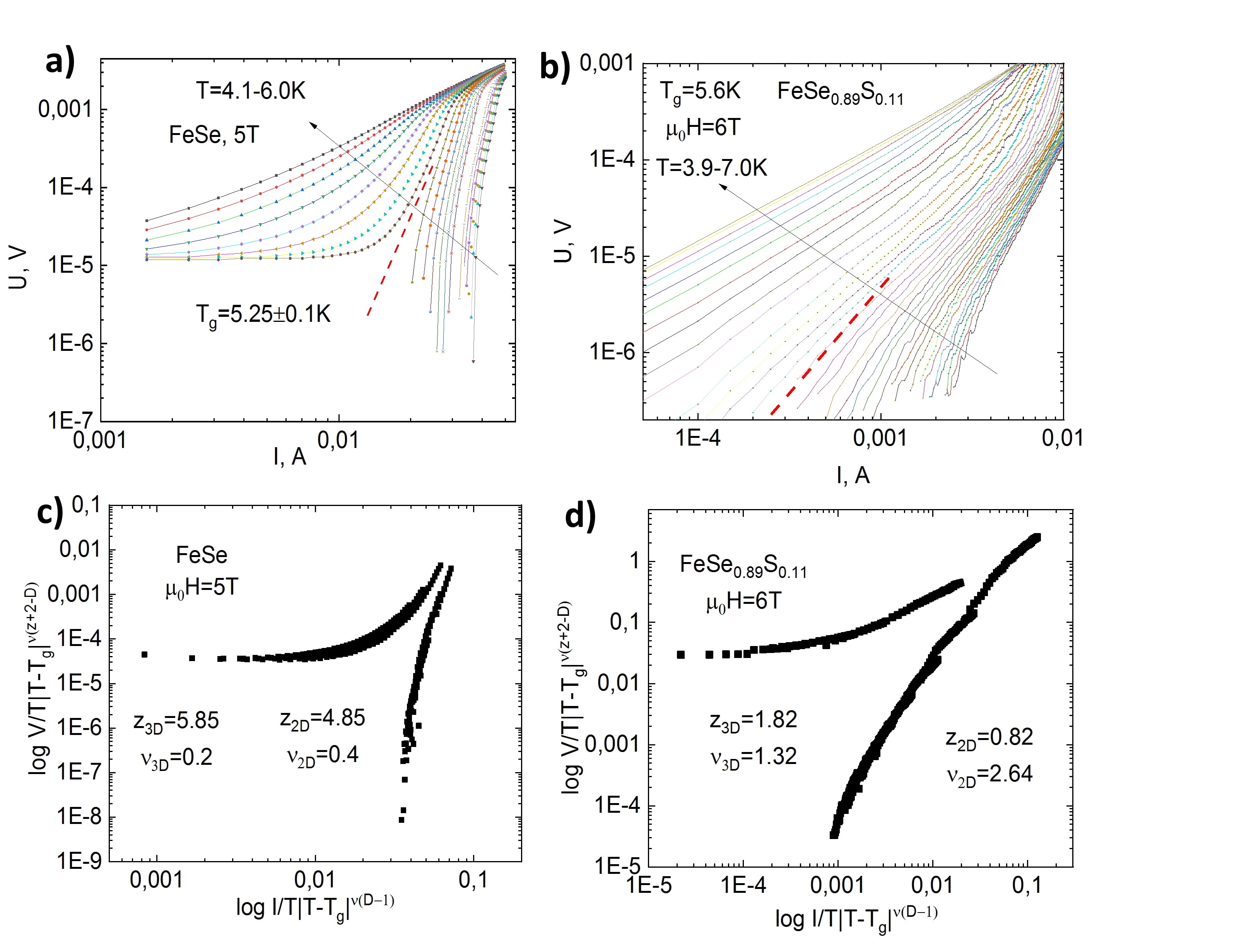}
	\caption{ a-b) The I-V characteristics of FeSe sample  and FeSe$_{0.89}$S$_{0.11}$ sample, with increment 0.1 K. c-d): The vortex-glass scaling for the I-V curves of FeSe sample, with determined $\nu$ and z fitting parameters. Insert: The FeSe$_{1-x}$S$_{x}$ bridges after FIB milling. Scale bar is 200 $\mu$m (left) and 50 $\mu$m (right). }
\end{figure}

Additionally, we checked the phase transition from the vortex-glass to vortex-liquid in FeSe (x=0 and 0.11) by using the I-V transport on micro bridges cut from single crystals as it shown in the inset of Fig.7(a). It should be noted that we considered the data only in the low dissipation regime ($E<10^{-4}$  V/cm). First, we evaluated  $T_g$ from the log(V)-log(I) curves, when the curvature changed from upturn to downturn character \cite{Koch1}, as shown by dashed line in Fig.7(a) for FeSe single crystal. As one can see, the value of $T_g$ at 5T is about $5.25\pm 0.1K$. According to the VG theory, the I-V curves near the transition temperature $T_g$  should diverge in two different branches by the scaling law. Therefore we plot the scaling $(V/I)(1-T/T_g)^{-\nu(z+2-D)}$ vs. $I/(T[1-T/T_g]^{2\nu}$ in the Fig.7(c-d), where $z,\nu$ are statical and dynamical exponents, $D$ is the dimensionality of the sample. According to theory \cite{Blatter},  3D vortex glass is a universal class with $s$ value in the range of $s=2.7-8.5$. If we assume D=3 for FeSe sample, the obtained critical exponent values are far beyond the VG model predictions. The quasi-2D (D=2) approach provides us the best scaling with the values  $\nu$=0.4 and z=4.85 (H=5T), and  the recalculated critical exponent value  $s=1.9\pm0.3$ for FeSe sample (where  $s=\nu(z+2-D)$) is also inconsistent with the assumption of 3D VG behavior. For FeSe$_{0.89}$S$_{0.11}$ sample the best scaling gives  the values $\nu$=2.64 and z=0.82 (H=6T), and   $s=2.2\pm0.4$. Here, the observed rather low $s$ exponent values  in all considered cases indicate a quasi-2D like behavior of VG phase. The non-3D vortex liquid-solid phase transition was also revealed in SmFeAsO$_{0.85}$\cite{Lee}  and possibly in  BaFe$_2$(As$_{0.68}$P$_{0.32}$) \cite{S Salem-Sugui Jr2} iron based superconductors.

\section{Upper critical field and phase diagram}

Using the results from R(T,H) measurements we obtained the upper critical fields temperature dependence with  H parallel to the $c$ axis as it shown in Fig.8(a-d). The H$_{c1}$  data obtained by us earlier \cite{Abdel}, makes it possible to estimate the Ginzburg-Landau parameter  
$ \kappa(0) =\lambda(0) /\xi(0) $ values using the formula: $\frac{{{2H}_{c1}}}{{{H}_{c2}}}=\frac{\ln \kappa +0.5}{{{\kappa }^{2}}}$. The obtained values were 79, 72, 64  for FeSe$_{1-x}$S$_x$  with x = 0, 0.04, 0.09, respectively. The estimated $ \kappa(0)$ for FeSe is close to value 72.3 reported in ref.  \cite{Hechang}. The thermodynamic upper critical field $H_c$ values at T=0K, calculated according to the expression $H_{c2}(0) = \sqrt{2}\kappa(0)H_c(0)$, are $H_c(0)$=0.132(4), 0.171(8), 0.204(3)T for the S substitution levels x = 0, 0.04, 0.09, respectively.

The majority of IBS are generally considered as multiple-gap superconductors, in particular, a two-band model can fit the  $H_{c2}(T)$ experimental data better than the single band WHH theory \cite{WHH}. The two-band model in the dirty limit proposed by Gurevich \cite{Gurevich} for $H_{c2}(T)$, which
takes into consideration orbital pair breaking effect refers to the following form: $a_1(\ln⁡t+U(h))+a_2(\ln⁡t+U(\eta h))+a_0(\ln⁡t+U(h))(\ln⁡t+U(\eta h))=0 $, where $t=T/T_c$-normalized temperature, $h=D_1\hbar H/2\phi_0 k_B T$ -normalized magnetic field, 

$a_1=1+\lambda_-/\lambda_0 $, $a_2=1-\lambda_-/\lambda_0$, $a_0=2\omega/\lambda_0$, $\lambda_-=\lambda_{11}-\lambda_{22},	\lambda_0= \sqrt{\lambda_-^2+4\lambda_{12} \lambda_{21}}$,	$\omega=\lambda_{11}\lambda_{22}-\lambda_{12}\lambda_{21}$, $\lambda_{ij}$  is the coupling constants matrix, $\eta=D_2/D_1$-the diffusion ratio. The function $U(x)$ has the following form: $U(x) = \psi(x+1/2)-\psi(x)$, where $\psi(x)$ is the digamma function. For our estimates, we use the intraband coupling constants $\lambda_{11}$ and $\lambda_{22}$ calculated from the $\mu$SR experiment \cite{Khasanov}. Assuming temperature dependence of the chemical pressure equivalent to external pressure, we interpolate the $\lambda_{11}$ coupling constant by linear function to obtain more precise $\lambda_{11}$ and $\lambda_{22}$  values. Also we suppose $\lambda_{12}$ = $\lambda_{21}$, in order to reduce the number of free parameters. The best fit of H$_{c2}^c$(T) data with the two-band model is presented in Fig.8(a-d). For comparison, we have also  fit our data with the single-band WHH model, as shown by green solid lines in Fig.8(a-d). The best fit parameters for both models are presented in Table I. One can see the electronic diffusivity D$_1$ varies a bit from 1.05 to 0.86, and the diffusivity ratio $\eta$ = D$_2$/D$_1$ slightly grows from 0.143 to 0.18 upon isovalent doping. The resulting $\eta$ values are slightly less than $\eta$ values obtained in ref.\cite{Lei1}. The observed insignificant increase in the diffusivity ratio $\eta$ suggests stable electronic mobility or an even change of the scattering rate for each band. Therefore, the S doping up to 0.11 does not really change the effective band structure of FeSe$_{1-x}$S$_x$ superconductor.  

\begin{table}[]
	\caption{\label{} Compilation of the FeSe$_{1-x}$S$_x$ samples parameters for the upper critical field.}
	\renewcommand{\arraystretch}{1.3}
	\begin{center}	 
		\begin{tabular}{|c||c||c||c|c||c||c|}
			\hline
			x & T$_c(50\%)$ (K)   & -dH$_{c2}^c$/dT (T/K)   & H$_{c2}^c$(0) (T) *
			& D$_1$(cm$^2$/s) & $\eta$ ($D_2/D_1$) & H$_{c2}^c$(0) (T)**\\ 
			\hline 
			0  & 9.36  & 1.92 &12.2  &1.05 &0.143 &17.0 \\ 
			\hline
			0.04 & 10.34 & 1.99 &13.9 &0.98 &0.143 &19.1\\ 
			\hline
			0.09 & 10.20 &2.19  &15.2  &0.86 &0.151 &20.8\\ 
			\hline
			0.11 & 10.09 &2.06  &14.3  &0.89 &0.180 &18.8\\ 
			\hline
			
		\end{tabular}
	\end{center}
	* - WHH model estimation, ** - two band model estimation.
\end{table}

\begin{figure}
	\includegraphics[width=1\textwidth]{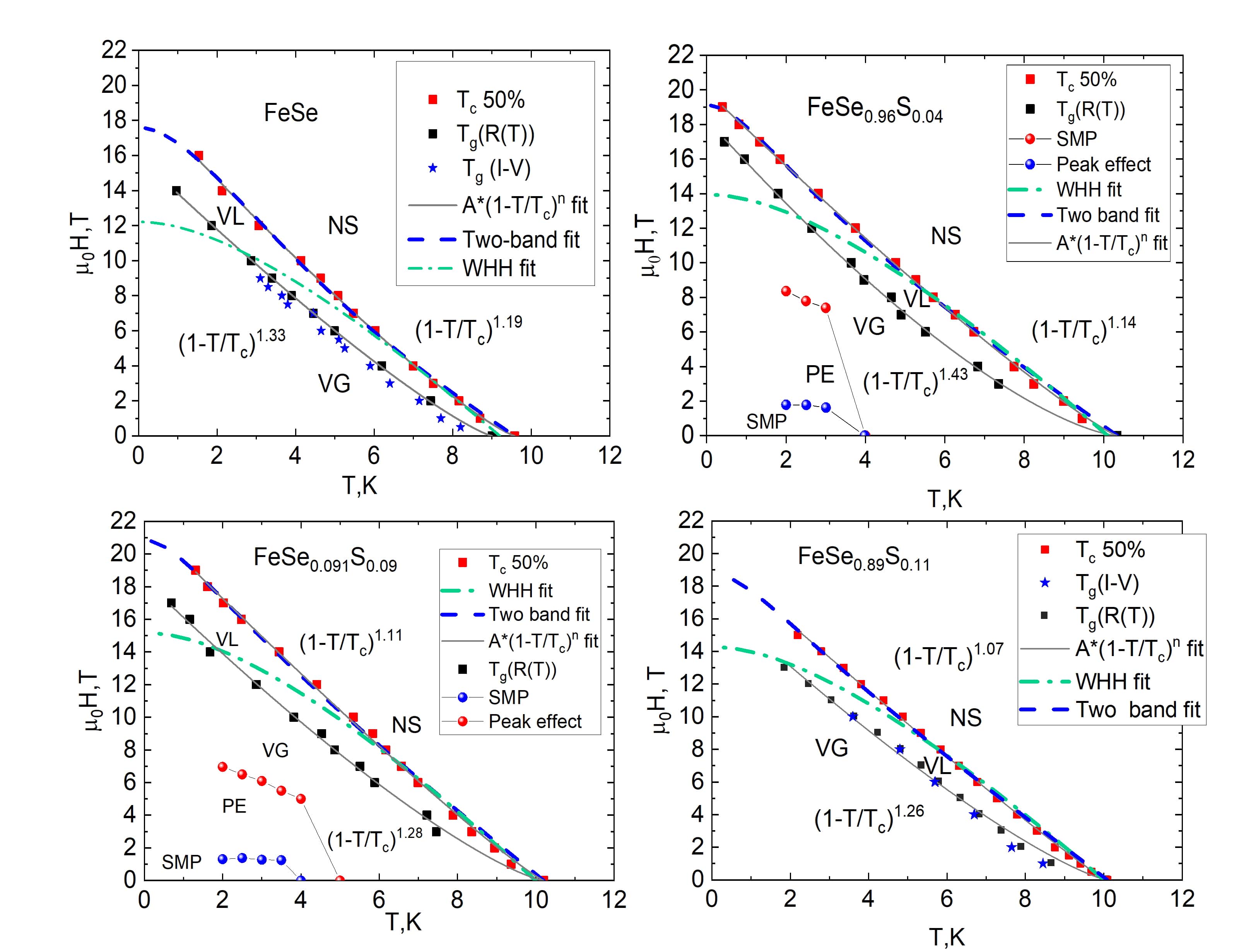}
	\caption{The FeSe$_{1-x}$S$_x$  (x=0, 0.04, 0.09, 0.11) phase diagram H vs. T with H$\parallel $c axis. The solid red lines  fit the experimental data with functional form $H(T) =H(0)\times(1 - T/T_c)^n$. The dashed blue lines fit the experimental data with the two band model. The designations are VG-vortex glass, NS- normal state,  VL - vortex liquid phase lines, SMP-second magnetization peak. The FeSe and FeSe$_{0.89}$S$_{0.11}$  data are obtained using the  R(T) and I(V) measurements.}
\end{figure}

Summarizing our data, we built the vortex phase diagram for the FeSe$_{1-x}$S$_x$ superconductor with various doping level. In Fig.8(a-d) we present the vortex glass, second magnetization peak, peak effect and upper critical field phase lines. We add the data, obtained in the I-V experiments, shown with blue stars, as well. The VG-VL line of the I-V measurements agrees well with the data obtained in the R(T) measurements. The experimental points defining the vortex glass to vortex liquid phase transition, and those corresponding to the upper critical field can be fitted nicely with the empirical formula $H(T) =H(0)\times(1 -T/T_c)^n$. The vortex liquid to vortex glass curves give n = 1.33 for FeSe, n = 1.37 for  FeSe$_{0.96}$S$_{0.04}$, n = 1.12 for FeSe$_{0.91}$S$_{0.09}$ and  n = 1.26 for FeSe$_{0.89}$S$_{0.11}$. For the upper critical field H$_{c2}$, n = 1.19, 1.14, 1.2, and 1.07  respectively. Similar values of $n$ exponent were shown for the Fe(Se,S) single crystals with low sulphur concentration elsewhere\cite{Aifeng Wang}.

\section{Conclusions}
In conclusion, we have studied the factors affecting vortex pinning in FeSe$_{1-x}$S$_x$ system. Our data evidences a complex interplay between various pinning contributions. We found the strong pinning arising from crystal structure distortions that plays an essential role in overall vortex pinning. The second magnetization peak was observed in all the studied samples, regardless of sulphur concentration.  We unambiguously showed that SMP originates from the transformation in the vortex lattice. Our studies point that peak effect correlates with S concentration , possibly caused by the $H_{c2}$ inhomogeneity in high fields, but it's nature is still debatable. 

We show the phase transition from vortex liquid to vortex glass for all the samples.  However, in FeSe$_{1-x}$S$_x$ up to x=0.11 the critical $s$-exponent is much less that predicted for the 3D case, thus supporting the existence of a quasi-2D-like phase transition. Our H$_{c2}^c$(T) data fits well by two-band model and show the rather stable electronic mobility or an even change of the scattering rate for each band with sulphur doping up to 0.11.  Summarizing the data obtained by several techniques, we present the detailed vortex phase diagram  for the FeSe$_{1-x}$S$_x$ compounds. 

\section{Acknowledgments}
VAV, AVS, TAR, SYG, AVD, OAS, BIM, EIM, and TEK acknowledge support from Russian Foundation for Basic Research Grant (no. 19-02-00888). The work of DAC and ANV is supported by Russian Foundation for Basic Research Grants 20-02-00561,  17-29-10007, by the program 211 of the Russian Federation Government, agreement No. 02.A03.21.0006, and by the Russian Government Program of Competitive Growth of Kazan Federal University. The measurements were done using research equipment of the Shared Facilities Center at LPI.

\end{document}